\documentclass[10pt,a4paper]{article}
\pdfoutput=1
\usepackage{amssymb}
\usepackage{amsthm}
\usepackage{amsfonts}
\usepackage{graphicx} 

 \newtheorem {theorem} {Theorem} 

 \advance \textheight by \topskip
\title{Modeling the prevalence of Schistosoma mansoni infection in an endemic population}

\author{
R. Oliveira-Prado \footnote{Escola de Enfermagem, Universidade Federal de Minas Gerais, Brazil} ,
M. Alvares de Souza Cabral\footnote{Faculdade de Medicina, Universidade Federal de Minas Gerais, Brazil} , \\
S. Oliffson Kamphorst\footnote{Departamento de Matem\'atica, Instituto de Ci\^encias Exatas, Universidade Federal de Minas Gerais, Brazil} ,
S. Pinto-de-Carvalho$^\ddagger$ ,\\
R. Corr\^ea-Oliveira\footnote{Centro de Pesquisas Ren\'e Rachou, Funda\c c\~ao Oswaldo Cruz, Brazil} ,
A. Gazzinelli$^*$ 
}
\date{}

\begin{document}
\maketitle

\begin{abstract}
We investigate theoretically if treatment alone can reduce the schistosomiasis's prevalence in an infected population, in a long-lasting sustainable way.  We use a non-linear system of ordinary differential equations  (a SI system combined with a logistic population growth) which describes the time evolution of  the non-infected and infected  populations, in terms of the recovering, infection, and  demographic rates. 

Our model leads to the conclusion that the only way to eliminate this endemic disease is to implement public health policies aimed at both treatment and environment. 

We apply our model to the endemic area of  Virgem das Gra\c cas, in Brazil, where the prevalence of Schistosiamisis in 2001 was greater than 50\%.
The epidemiological data are extracted from a longitudinal study carried on the region between 2001 and 2010 and the demographic parameter from official Brazilian population data. When these estimated parameters  are entered, our model gives a  limit prevalence for Virgem das Gra\c cas of 11\%, which is still significantly high even though treatment in accordance to the government regulations is systematically performed.
This estimative reinforces once again, that in order to eliminate Schistomiasis, public health policies aimed at treatment, sanitation, including snail control, and health education programs are mandatory.

\end{abstract}

%\keywords{schistosomiasis mansoni \and treatment \and non linear modeling of prevalence }

\section{Introduction}
Schistosomiasis is a major neglected tropical disease, acquired when people get in contact with water infested with the larval form of the parasite, during routine agricultural, domestic, occupational and recreational activities (Steinmann et al 2006).
The global importance of schistosomiasis is significantly high, as over 200 million people are infected worldwide and close to 800 million are at risk (King and Dangerfiel-Cha 2008).
In Brazil, schistosomiasis is endemic in 9 states, affecting almost 6 million people. 
It is intimately associated with poor socioeconomic conditions, and a vicious cycle takes place: as the lack of sanitation and treated water increases the risk of infection, the increased disability, related to the impact of chronic and recurrent infection, reduces productivity and perpetuates poverty (King 2011).

The World Health Organization (WHO) has launched a goal challenge to globally eliminate schistosomiasis by the year 2025 and has indicated that preventive chemotherapy with Praziquantel is essential as well as complementary public-health interventions are strongly recommended to achieve elimination and interruption of transmission (WHO 2013).
Praziquantel was synthesized and tested against schistosomiasis and cestodes in the mid-70's (Andrews et al 1983) and is effective against all species of these parasites that affect humans: {\sl Schistosoma mansoni} (which is the only species found in Brazil), {\sl S. hematobium, S. japonicum, S. intercalatum} and  {\sl S. menkongi}. 
However, there is a growing evidence that treatment alone is not sufficient to achieve the goal of the program (Ross et al 2015). 

The Brazilian government's proposal to reduce the transmission of  schistosomiasis
focuses on reducing the prevalence of the disease through periodic treatment  of the population together with a more comprehensive approach that should include  water treatment, adequate sanitation and snail control (Brasil 2014). Despite these efforts, Brazil is still the country  that  concentrates  the greatest  number of  schistosomiasis cases in South America (WHO 2010).

Schistosomiasis is a complex disease and clearly, as more knowledge is gained about its biology, epidemiology, and transmission,  more effective methods of intervention can be developed. 
To improve its comprehension, statistical methods have been proposed to analyze the treatment efficacy on transmission and control. However, only a few studies use longitudinally collected data as presented here and most of them concerns Africa.
Besides that, many theoretical studies focus on cure and reinfection rates and do not pay significant attention to its dynamics.

Since the pioneering work of Bernoulli on measles in the XVIII century (Bacaer 2011), 
qualitative mathematical modeling has proved itself to be also a valuable tool in the prediction of epidemic trends and the design of control programs  for many infectious diseases (see, for instance, Anderson and Nokes 1991;  Garnett and Anderson 1995; Barbour 1996), allowing the investigation to go beyond the other methods, to extrapolate the data for many years ahead and to simulate how the scenario will be, based on the processes that have already occurred. 

In the present paper, our goal is to investigate theoretically if treatment alone can reduce the prevalence in an infected population, in a long-lasting sustainable way.  To do so, we use a non-linear system of ordinary differential equations (more precisely, a SI system combined with a logistic population growth) which describes the time evolution of  the non-infected and infected  population, in terms of the recovering, infection, and demographic rates. 

Ordinary and partial differential equations have long been used to model schistosomiasis infection in a population.
The earlier models introduced by Macdonald (1965) and Hairston (1962) described the dynamics of transmission between man, mammals and snails. 
Since that, many articles were devoted to those models 
improving the differential equations and looking at control strategies (see, for instance, Woolhouse 1991, 1992; Barbour 1994; Medley and Bundy 1996; William et al 2002) . 
More recently, linear SI-models dividing the human population in strata defined by age, and including snails and/or bovines (Gurarie et al 2010; Xiang et al 2013; Zhao and Milner 2008 ) were developed, leading to very large systems with many parameters, which may be very difficult to estimate from real data.
None of these models, however,  manages to reproduce in a realistic way the dynamics of prevalence reduction after treatment, using a cohort study. A good exception is the work of French et al (2010) that estimates, via partial differential equations, the effect of  multiple treatment rounds  in the reduction of environmental transmission.

The analysis of our model shows that although regular treatment could substantially reduce the prevalence, it is not enough by itself to clear prevalence in a population. 

In fact, previous studies  have already shown that treatment alone leads to lower levels of infection, and a decrease in morbidity and mortality. However,  prevalence can return to the previous levels in a short period of time due to the reinfection process (McManus and Loukas 2008). So, as long as (re)infection occurs, there is no expectation of disease elimination. 

Our model leads to the conclusion that the only way to eliminate this endemic disease is to implement public health policies aimed at both treatment and sanitation. The results are also supported by recent reports suggesting that treatment has not been as efficacious as expected even when repeatedly applied in schistosomiasis endemic regions (Secor 2015).

We apply our model to an endemic zone in Brazil, where the prevalence in 2001 was greater than 50\%. 
The epidemiological data are extracted from a longitudinal study carried on the region between 2001 and 2010 and the demographic parameter from official Brazilian population data, as described on section \ref{virgem}.  
When these estimated parameters  are entered, our model gives a  limit prevalence of 11\%, which is still significantly high even though treatment in accordance to the government regulations is systematically considered.
We conclude that if health measures in conjunction with environmental and educational interventions are not taken it will be impossible to achieve elimination and interruption of transmission of the disease.

\section{The mathematical model}

We consider two variables: $p_1(t)$, which stands for the number of infected individuals at time $t$,  and $p_0(t)$, the number of non-infected individuals at time $t$. The total population at an instant $t$ is then $P(t)= p_0(t)+p_1(t)$. 

We make the following assumptions:
\begin{itemize}
\item the evolution of the total population follows a logistic equation (see, for instance, Murray 2001) with a maximum capacity threshold; 
\item  birth, mortality and migration rates  are identical for infected and non infected populations; 
\item the infection rate is constant and uniform and does not depend on age, occupation or any other factor, except for newborns that are free of infection;
\item treatment is applied to infected individuals continuously at a constant rate; 
\item once effectively treated any individual will be free of infection unless he/she is reinfected.
\end{itemize}

\begin{figure}
\begin{center}
\includegraphics[angle=270,width=.4\hsize]{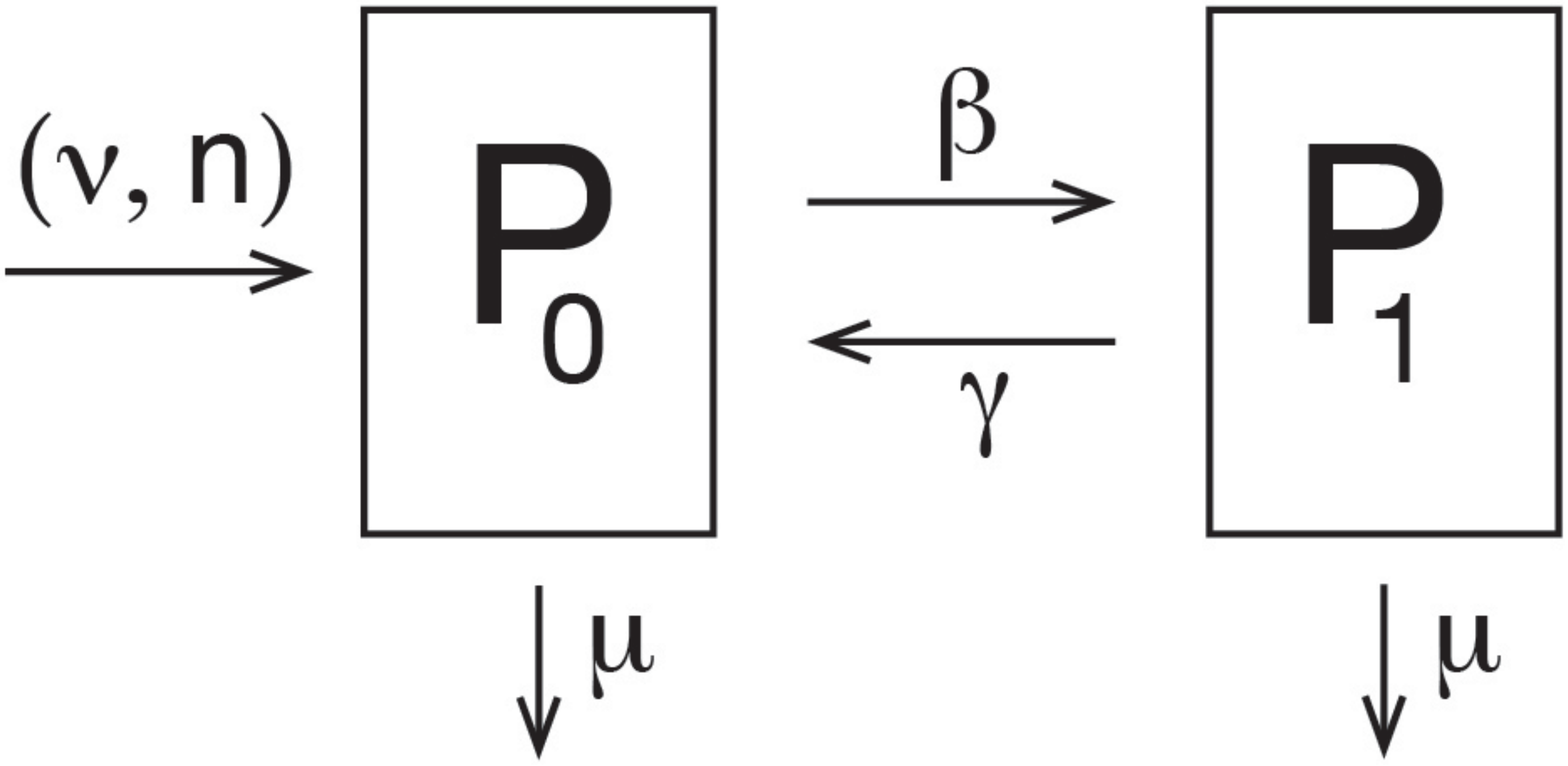}
\end{center}
\caption{Schematic model}\label{fig:esquema}
\end{figure}

Our model is schematically shown on Figure~\ref{fig:esquema} and is represented by the following system of differential equations
\begin{eqnarray}
\frac{dp_0}{dt}&= &-\beta p_0-\mu p_0+\gamma p_1+\nu(n-p_0-p_1)(p_0+p_1)\label{eq:sistema} \\
 \frac{dp_1}{dt}&=& \beta p_0-\mu p_1-\gamma p_1\nonumber
  \end{eqnarray}
  defined on the set $\Delta=\{(p_0, p_1), 0\leq p_0+p_1\leq n\}$.  

The birth rate $\nu$, the withdraw rate $\mu$, the infection rate $\beta$ and the maximum population size $n$ are strictly positive parameters, but the recovering rate $ \gamma$ is only positive, since we may consider no treatment at all, which corresponds to $\gamma=0$.

\begin{table}[h]
\begin{center}
\begin{tabular}{|c|l|}
\hline
$p_0 $ & non-infected individuals \\ \hline
 $ p_1 $ & infected individuals \\ \hline
$t$ & time (years) \\ \hline
 $  \nu $ & birth rate parameter\\ \hline
 $ \mu$ & withdraw (death or migration) rate \\ \hline
 $ n$& maximum population size\\ \hline
$  \beta$ & infection rate  \\ \hline
$\gamma$ & recovering rate\\ \hline
\end{tabular}
\end{center}
\caption{List of symbols}
\label{tab:symbols}
\end{table}

Summing up the two lines of system (\ref{eq:sistema}) we note that the total population $P= p_0 + p_1$ satisfies the logistic differential equation
\begin{equation}
\frac{d P}{dt} = \nu\left(\frac{\nu n-\mu}{\nu}-P\right)P \nonumber
\label{eq:logistica}
\end{equation}
where $\displaystyle \frac{\nu n-\mu}{\nu}\leq n$ is the the carrying capacity of the population and so we must have $\nu n-\mu>0$.

\begin{figure}
\begin{center}
\includegraphics[width=.5\hsize]{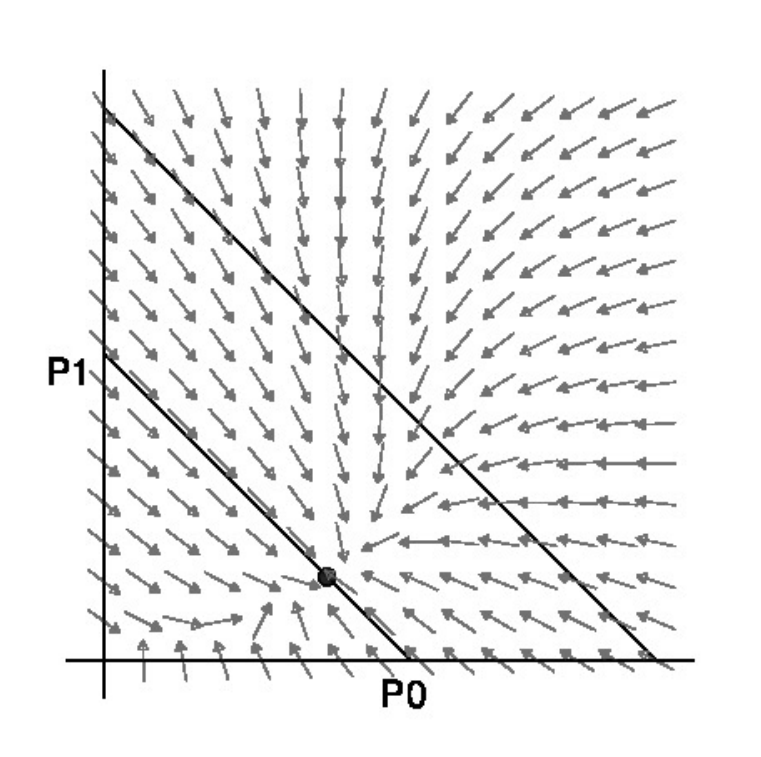}
\end{center}
\caption{$\Delta$ is invariant and $(p_0^\star,p_1^\star)$ is an attractor.}\label{fig:invariante}
\end{figure}

The set $\Delta=\{(p_0, p_1), 0\leq p_0+p_1\leq n\}$ is invariant by the flow of the system (\ref{eq:sistema}) since 
$$
\begin{array}{lcll}
\displaystyle  < \left. \left(\frac{dp_0}{dt},\frac{dp_1}{dt}\right) \right|_{(0,p_1)},(1,0) > & = &  \left(\gamma+\nu (n-p_1)\right)p_1>0 & \hbox{ for } p_1 > 0\\
\displaystyle   < \left. \left(\frac{dp_0}{dt},\frac{dp_1}{dt}\right) \right|_{(p_0,0)},(0,1) > & = & \beta p_0>0 &  \hbox{ for  }  p_0 >  0\\
\displaystyle   < \left. \left(\frac{dp_0}{dt},\frac{dp_1}{dt}\right) \right|_{\{p_0+p_1=n\}},(1,1) > & = & -\mu n<0 &
\end{array}
$$
which shows that the vector field defined by  (\ref{eq:sistema}) points inward $\Delta$, except at $(0,0)$, as illustrated on Figure \ref{fig:invariante}.

System (\ref{eq:sistema}) has two equilibria:  the origin $(0,0)$ and  the point
$\left(p_0^\star, p_1^\star\right)$  where
$$
 p_0^\star=\frac{ (\nu n -\mu)(\gamma +\mu)}{\nu(\gamma+\beta+\mu)} \ \  \mbox{and}\ \ 
 p_1^\star= \frac{\beta}{\gamma+\mu} p_0^\star
$$  
As the parameters $\nu, \mu, \beta > 0 $, $\gamma\geq 0$ and $\nu n-\mu>0$,  it is clear that $p_0^{\star}$ and $p_1^{\star}$ are both positive. Moreover 
the total population for this equilibrium is $p_0^{\star} + p_1^{\star} = n - \mu/\nu$  (the carrying capacity)  and is smaller than $n$. This implies that this non trivial equilibrium point in inside $\Delta$.

The point $(0,0)$ is a saddle with eigenvalues $\lambda_+ =\nu n-\mu>0$  and $\lambda_-=-(\gamma+\beta+\mu)<0$. The eigenvector associated to the positive eigenvalue  is  $\vec v_+=(-1,1)$ and so its stable direction is  outside $\Delta$. Moreover, as the flow is transverse to both axis, the stable curve of $(0,0)$ must also stay outside $\Delta$.

On the other hand, the equilibrium point
 $(p_0^\star, p_1^\star)$ has eigenvalues $-(\mu+\gamma+\beta)$ and $-(\nu n-\mu)$  which are both negative, so it is a local attractor. 
Studying the behaviour of the flow on the lines $p_0+p_1=k$ (illustrated on Figure \ref{fig:invariante}) we have that
 $$< \left. \left( \frac{dp_0}{dt}, \frac{dp_1}{dt} \right) \right|_{\{p_0+p_1=k\}},(1,1) >=\frac{k(\nu n-\mu-k\nu)}{\nu}\quad
\mbox{is}\quad\left\{
\begin{array}{cc}
>0 &\mbox{if} \ \ k< n-\frac{\mu}{\nu}\\
=0 &\mbox{if}  \ \  k=n-\frac{\mu}{\nu}\\
<0 &\mbox{if} \ \ k>n-\frac{\mu}{\nu}
\end{array} \right.
$$
and so $(p_0^\star,p_1^\star)$ is actually a global attractor on $\Delta$. This means that any solution starting from an arbitrary point in $\Delta$, except for the origin $(0,0)$,  will approach $(p_0^\star,p_1^\star)$ as the time $t$ becomes very large. 

This translates into the following
\begin{theorem}
Regardless of the current values of the infected individuals, if $\nu,\gamma,\beta,\mu,n$ are positive parameters and $\nu n-\mu>0$, and if the total population is not zero and does not exceed $n$ then the prevalence satisfies
$$\frac{p_1(t)}{p_0(t)+p_1(t)}\longrightarrow\frac{\beta}{\gamma+\mu+\beta}\ \ \mbox{as}\ \ t\to\infty.$$
\end{theorem}

{Proof:} Given any non zero initial condition on $\Delta$, the solution $(p_0(t),p_1(t))$ will tend to the global attractor $ (p_0^\star,p_1^\star)$. Then the prevalence 
$$\frac{p_1(t)}{p_0(t)+p_1(t)}\longrightarrow\frac{p_1^\star}{p_0^\star+p_1^\star}=\frac{\beta}{\gamma+\mu+\beta}\ \ \mbox{as}\ \ t\to\infty.$$
\qed

Differentiating $\frac{\beta}{\gamma+\mu+\beta}$ with respect to $\gamma$ we get 
$$\frac{\partial}{\partial \gamma} \left( \frac{\beta}{\gamma+\mu+\beta} \right) = \frac{-\beta}{(\gamma+\mu+\beta)^2}<0 $$
 which show us that, for $\mu$ and $\beta$ fixed, the limit prevalence strictly decreases with the treatment rate $\gamma$.

On the hand, differentiating $\frac{\beta}{\gamma+\mu+\beta}$ with respect to $\beta$ we get 
$$\frac{\partial}{\partial \beta} \left( \frac{\beta}{\gamma+\mu+\beta} \right) = \frac{\gamma+\mu}{(\gamma+\mu+\beta)^2}>0 $$
 which shows that, for $\mu$ and $\gamma$ fixed, the limit prevalence strictly  increases with $\beta$.

So,  treatment can decrease the prevalence but  the only way to eliminate the disease is to avoid (re)infection ($\beta=0$). 

\section{Virgem das Gra\c cas's longitudinal study}\label{virgem}

\begin{figure}
\begin{center}
\includegraphics[width=.8\hsize]{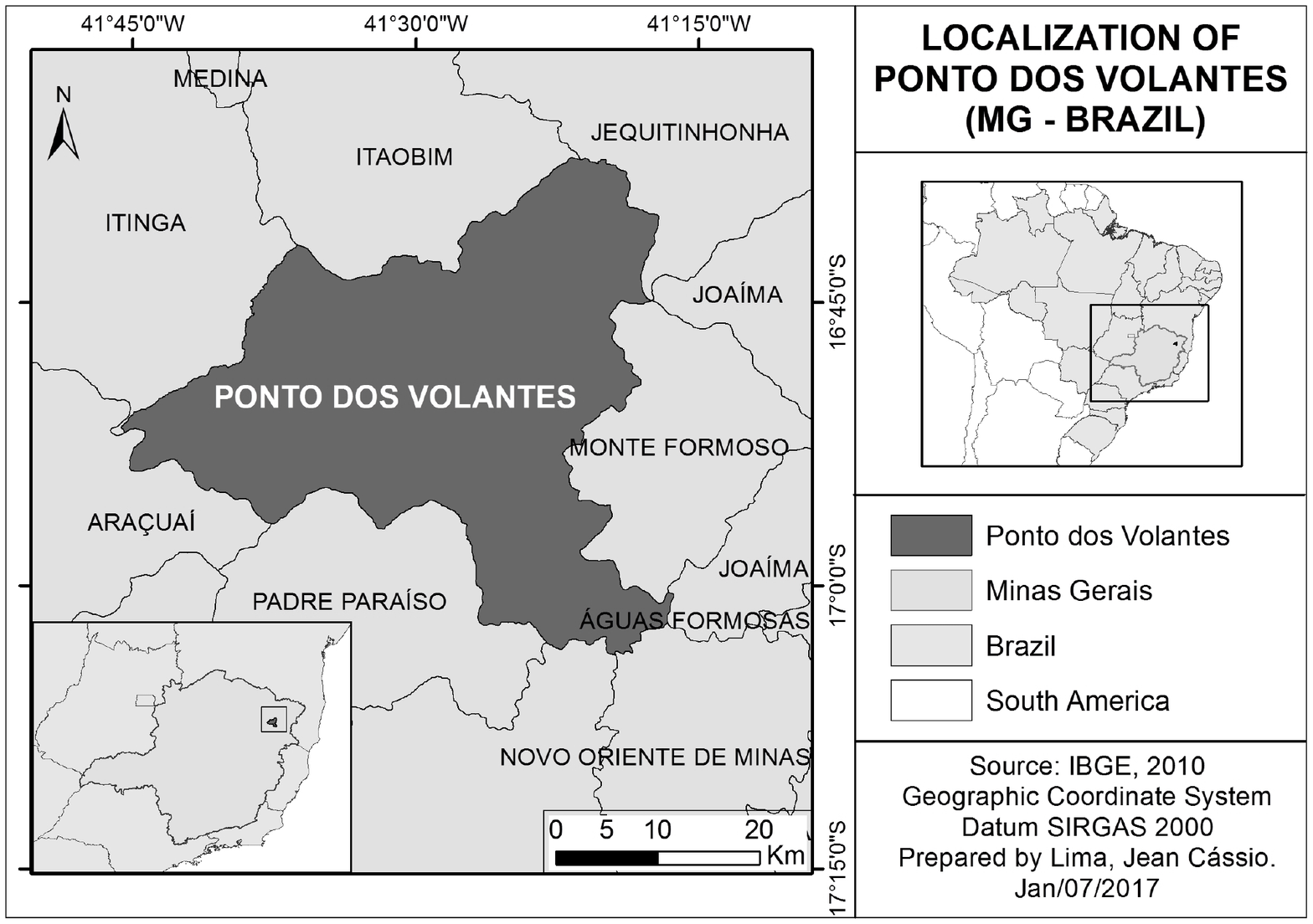}
\caption{Map of the municipality of Ponto dos Volantes, Minas Gerais State, Brazil }\label{fig:mapa}
\end{center}
\end{figure}

The parameters of our model were calibrated from the data collected by a longitudinal study performed between 2001 and 2010 with the total population from the area of  Virgem das Gra\c cas, a rural community located in the municipality of Ponto dos Volantes in northern Minas Gerais, Brazil (see Figure \ref{fig:mapa}).

The population lives in four dispersed hamlets (Cardoso 1, Cardoso 2, Cardoso 3  and Su\c cuarana) along the Cardoso and Su\c cuarana streams, and in the central village. The local economy is based on corn and manioc cultivation and cattle husbandry, for subsistence, and remittances from migrants working in cities. 

Parasitological examinations were carried out in 2001, 2002, 2005, 2009 and 2010, followed by treatment at each time point as reported by Matoso et al 2012. 

Stool samples from  all individuals who volunteered were examined for {\sl S. mansoni} eggs using the Kato-Katz method (Katz et al 1972). They received 3 plastic containers for fecal samples and were instructed to deposit one fecal sample per day into each container and return it immediately to the collecting point, where the samples were stored at $4^o C$. Two slides for each stool sample (a total of 6 slides per individual) were prepared for egg counting within 24 hours of collection, as described by Gazzinelli et al (2006). Infected individuals were identified by the presence of parasite eggs on the samples. The global data are presented in Table \ref{tab:dados} below.

\begin{table}[h]
\begin{center}
\begin{tabular}{|l|c|c|c|c|c|}\hline
 
 &  \bf 2001 & \bf 2002 & \bf 2005 & \bf 2009 &  \bf 2010 \\
 \hline
 \bf Tested
& 571 & 521 & 519 & 587 & 490 \\
\hline
 \bf Infected
& 323 & 87 & 121 & 155 & 64 \\
\hline
\end{tabular}
\end{center}
\caption{Infection data per time point:  total  number of individuals examined by Kato-Katz (tested--line 1) and total  number of positive individuals, ie having at least one egg in one of the slides (infected--line 2) }\label{tab:dados}
\end{table}

The treatment protocol used in this study was issued by the Ministry of Health in 1998 (Brasil 1998).
Its guideline recommends to treat individuals according to the measured percentage  of infection in the population (prevalence) at the time point. 
The stated recommendation was to treat only the infected (egg positive) individuals  in places where the prevalence was bellow 25\%.
For prevalence between 25\% and 50\%, treatment should also include house cohabitants. Finally, for a prevalence over 50\%, wide treatment of the entire population was recommended.
We remark that, as determined by the Brazilian health regulations, pregnant and lactating women could not receive treatment  (Brasil 2014).

Since the prevalence in 2001 (baseline) was over 50\%, all eligible participants were treated with a single oral dose of Praziquantel  of 50 mg/kg for those with age over 15 years and 60 mg/kg for those below, in agreement to the Ministry of Health regulations.  In the following years, individuals where treated  according to the observed prevalence.

\subsection{Estimating the parameters of the model}

\subsubsection{The withdraw rate $\mu$}

Although the municipality of Ponto dos Volantes is located in the state of Minas Gerais, on the southeast  of Brazil,  its social indicators are more similar to the ones of poorer states of the semi arid Northeast region (Gomes et al 2010).

To estimate the withdraw rate $\mu$,  composed by the mortality and the migration flux, we first observe that, according to the Brazilian Census of 2010 (IBGE 2010), the average number of death per 1000 habitants for states in the semi arid region runs from 4.43 to 6.26. Also according to  Gomes et al (2010) migrations have a small effect in the population evolution. 

So we will  take the average mortality rate for the semi arid region, $5.24/1000$, and round it up, to include the negative migration flux, ending up with the estimated value $\mu = 0.006$ for our population. 

\subsubsection{The recovering rate $\gamma$}

Our model assumes that treatment is applied in a continuous basis. 
The parameter $\gamma$ represents then, the relative rate at which infected individuals are converted to non infected (recovered), and is called the recovering rate. 

We assume that recovering is  exclusively due to treatment with Praziquantel. However, throughout the longitudinal study performed, not all the infected individuals were treated for different reasons, such as traveling at the time of treatment, being pregnant or breastfeeding or not being found at home after several visits.
In Table \ref{tab:gamma} we present the total number of infected individuals per year of study ($I$), and the number of those, among them, who received treatment ($T$).    

On the other hand,  we must also take into account  that people can remain infected even after treatment. 
According to the manufacturer of Praziquantel (Merck 2016)  this drug is effective against adult schistosomes, but not developing schistosomula, which are present early in infection. Andrews et al (1983) estimate its efficacy in 95$\%$. 
We estimate then that the number of recovered individuals corresponds to  95\% of the treated ones at each time point. 

\begin{table}[h]
\begin{center}
\begin{tabular}{|l|c|c|c|c|c|c|}\hline
 & & \bf 2001 & \bf  2002 & \bf  2005 & \bf  2009 & \bf  2010 \\
\hline \hline
\bf  Infected &$I$& 323 & 87 & 121 & 155 & 64 \\
\hline
 \bf Treated & $T$& $303$ & 69 & 81 & 145 & 55 \\
 \hline
 \bf Recovering & $\gamma$ &0.89 & 0.75&0.64& 0.89& 0.81\\
\hline
\end{tabular}
\end{center}
\caption{Treatment data}\label{tab:gamma}
\end{table}

So, for each year of the study, we approximate the recovering rate $\gamma$ by the fraction of recovered individuals 
$$\gamma\approx \frac{0.95 \times T}{I}$$
obtaining, after rounding up, the estimated  recovering rate for each year, showed on the last line of Table \ref{tab:gamma}.
Averaging the obtained values we consider $\gamma=0.8$.

\subsubsection{The infection rate $\beta$}

In order to estimate the infection rate from the data, we fix our attention on a group of individuals assumed to be cleared of infection at a given time point  (individuals with negative stool test and individuals with positive stool test who have received treatment) and who have been tested in the next time point. We call this the control group, and the number of individuals in each control group ($C$) is given in the first line of Table \ref{tab:beta}. We denote by $I$ the number of individuals tested and infected at each step, as listed in the second line of Table \ref{tab:beta}.

\begin{table}[h]
\begin{center}
\begin{tabular}{|l|c|c|c|c|c|}
\hline
  &  & \bf 2002 &  \bf 2005 &  \bf 2009 &  \bf 2010 \\
\hline \hline
 \bf   Tested & $C$ &  465 & 431 & 398 & 449 \\
\hline
   \bf Tested and infected &$I$ &   69 & 101 & 101 & 52 \\
\hline
  \bf  Mean time between tests (years) &$\Delta t$ &  1.38 & 3.10 & 3.60 & 1.37   \\
\hline
 \bf   Infection rate &$\beta$ &   0.12  & 0.09  & 0.08 & 0.09\\
\hline
\end{tabular}
\end{center}
\caption{Infection data}\label{tab:beta}
\end{table}

For example, the data in the first column show that 465 individuals were tested negative or treated in 2001 and were tested again in 2002. From those, 69 tested positive at the time point 2002. The mean elapsed time between the old and the new tests was in fact 1.38 years, slightly above one year.

We suppose in our model that infection takes place in a continuous manner, implying  an exponential increase on the number of infected individuals (or equivalently, an exponential decrease on the number of non infected individuals).
Therefore the fraction of individuals remaining non infected after a time $\Delta t$ is  given by
$$\frac{C-I}{C}=e^{-\beta\Delta t}\quad\quad \mbox{or}\quad\quad \beta=-ln\left(1-\frac{I}{C}\right)\frac{1}{\Delta t}.$$

The computation of $\beta$ between each time point is shown on the last line of Table \ref{tab:beta}.
After averaging  these values and rounding up  we end up with the estimated value of $ \beta=0.1$.
 
\subsubsection{ Remarks about the model and the estimation of the parameters}

Models are idealized. Nonetheless, if the preliminary hypothesis are correct, the qualitative information they provide are valid. 

The major difficulty when applying models to specific situations is to obtain the parameters. In our specific case, it begins with the characterization of infected or non-infected persons by egg counting on stool samples. As we consider infected any number of discovered eggs, the experimental error is minimized, but it is not null.

For the estimation of the recovering rate we assumed that, during treatment, there is no reinfection, which certainly is not completely true. In practice,  treatment is not applied in a continuous manner.

For the estimation of the infection rate we have assumed that it does not depend on seasons and rain, neither on the age of the individuals or occupation, nor on specific immunological defenses. 

In all cases, we took averages, which certainly do not take into account the individual variations.

Finally, there is no population census specific of Virgem das Gra\c cas, and we took estimations from biggest regions.

So we must look to the numerical conclusions as illustrative numbers and not as exact ones.

\subsection{The limit prevalence}

With our estimation of the parameters  we can then conclude that the prevalence will approach the
$$\mbox{limit prevalence}=\frac{\beta}{\gamma+\beta+\mu}=\frac{0.1}{0.8+0.1+0.006}\approx 0.11$$
meaning that  a strategy of treatment alone will lead, in some  years to a prevalence around 11\%.
 
\section{Conclusion}

In this work we have answered negatively to the question if mass treatment with Praziquantel applied to endemic communities could lead to important and long-lasting sustained reduction of {\sl S. mansoni} reinfections and prevalence of infection independently of other interventions.

We have  estimated the limit prevalence for Virgem das Gra\c cas as 11\%, which is in agreement with the estimates of the Brazilian Ministry of Health studies for the region (5 to 15\%) (Brasil 2016).

We can also observe that even if  treatment were given to 100\% of the infected individuals, we would have $\gamma=0.95$ and the limit prevalence will be of the order of 9\%. With no treatment at all ($\gamma=0$)  it will tend to 95\%, meaning that in the limit, only very young children would not be infected.

Our model predicts, so, that the only way to eliminate the endemic disease is to avoid (re)infection .  Since the recovering rate only changes the speed of approaching the equilibrium, then, alone, treatment can not eliminate the disease, although reducing significantly its prevalence. 

In other words, our conclusion is that, in order to eliminate Schistomiasis, public health policies aimed at treatment, sanitation, including snail control, and health education programs are mandatory.

\vskip.8cm
{\bf Acknowledgements}: 
The authors thank Tropical Medicine Research Center, National Institutes of Health-TMRC-NIH, USA, (Number 1P50AI098507-01), Funda\c c\~ao de Amparo \`a Pesquisa de Minas Gerais - FAPEMIG, Conselho Nacional de Desenvolvimento Cient\'\i fico e Tecnol\'ogico - CNPq, and Instituto de Ci\^encia e Tecnologia em Doen\c cas Tropicais - INCT/DT .
  ROP was funded by the Programa Nacional de P\'os-Doutorado/Coordena\c c\~ao de Aperfei\c coamento de Pessoal de N\'\i vel Superior - PNPD/CAPES.

\end{document}